\documentclass[traditabstract,longauth,letter]{aa} % for the long lists of affiliations
\usepackage{graphicx, txfonts}
\usepackage{threeparttable}
\newcommand{\less}{\raisebox{-1.1mm}{$\stackrel{<}{\sim}$}} 
 
\newcommand{\msol}{\mbox{M$_{\odot}$}} 
 
\newcommand{\msolyr}{{M$_{\odot}$}\,yr$^{-1}$} 
\newcommand{\mdot}{$\dot{M}$}

\newcommand{\ks}{km s$^{-1}$} 
 
\begin{document}
\title{Herschel PACS and SPIRE imaging of CW Leo
\thanks{Herschel is an ESA space observatory with science instruments
provided by European-led Principal Investigator consortia and with important participation from NASA.}
}

\author{
D. Ladjal\inst{1}
     \and
M.J. Barlow\inst{2}
     \and
M.A.T. Groenewegen\inst{3}
     \and
T. Ueta\inst{4}
     \and
J.A.D.L. Blommaert \inst{1}\and
M. Cohen\inst{5}\and
L. Decin\inst{1,6}\and
W. De Meester\inst{1}\and
K. Exter\inst{1}\and
W. K. Gear\inst{7}\and
H. L. Gomez\inst{7}\and
P. C. Hargrave\inst{7}\and
R. Huygen\inst{1}\and
R. J. Ivison\inst{8}\and
C. Jean\inst{1}\and
F. Kerschbaum\inst{9}\and
S. J. Leeks\inst{10}\and
T. L. Lim\inst{10}\and
G. Olofsson\inst{11}\and
E. Polehampton\inst{10,12}\and
T. Posch\inst{9}\and
S. Regibo\inst{1}\and
P. Royer\inst{1}\and
B. Sibthorpe\inst{8}\and
B. M. Swinyard\inst{10}\and
B. Vandenbussche\inst{1}\and
C. Waelkens\inst{1}\and
R. Wesson \inst{2}
}

%  \offprints{}

\institute{Instituut voor Sterrenkunde,
         Katholieke Universiteit Leuven, Celestijnenlaan 200D, B--3001 Leuven, Belgium
	      \email{Djazia.Ladjal@ster.kuleuven.be}
	      \and
           Department of Physics and Astronomy, University College London, 
           Gower Street, London WC1E 6BT
         \and
	     Koninklijke Sterrenwacht van Belgi\"e, Ringlaan 3, B--1180 Brussels, Belgium %3
         \and
Dept. of Physics and Astronomy, University of Denver, Mail Stop 6900, Denver, CO 80208, USA %4 Ueta
\and
Radio Astronomy Laboratory, University of California at Berkeley, CA 94720, USA%4->5
\and
Sterrenkundig Instituut Anton Pannekoek, Universiteit van Amsterdam, %5->6
Kruislaan 403, NL--1098 Amsterdam, The Netherlands
\and
School of Physics and Astronomy, Cardiff University, 5 The Parade, Cardiff, Wales CF24 3YB, UK %6->7
\and
UK Astronomy Technology Centre, Royal Observatory Edinburgh, Blackford Hill, Edinburgh EH9 3HJ, UK %7->8
\and
University Vienna, Department of Astronomy, T\"urkenschanzstrasse 17, A--1180 Wien, Austria%8->9
\and
Space Science and Technology Department, Rutherford Appleton Laboratory, Oxfordshire, OX11 0QX, UK %9->10
\and
Dept of Astronomy, Stockholm University, AlbaNova University Center, 
Roslagstullsbacken 21, 10691 Stockholm, Sweden %10->11
\and
Department of Physics, University of Lethbridge, Alberta, Canada %11->12
}

\date{Received:2010 / accepted:2010 }

\abstract{
Herschel PACS and SPIRE images have been obtained over a 30\arcmin$\times$30\arcmin\ area around the 
well-known carbon star CW Leo (IRC +10 216).
An extended structure is found in an incomplete arc of $\sim$22\arcmin\ diameter, which is cospatial with the termination 
shock due to interaction with the interstellar medium (ISM) as defined by Sahai \& Chronopoulos from ultraviolet GALEX images.
Fluxes are derived in the 70, 160, 250, 350, and 550 $\mu$m bands in the region where the interaction 
with the ISM takes place, 
and this can be fitted with a modified black body with a temperature of 25$\pm$3~K.
Using the published proper motion and radial velocity for the star, we derive a heliocentric space motion of 25.1 \ks. 
Using the PACS and SPIRE data and the analytical formula of the bow shock structure, we infer a
de-projected standoff distance of the bow shock of $R_{0} = (8.0 \pm 0.3) \times 10^{17}$ cm.
We also derive a relative velocity of the star with respect to the ISM of $(106.6 \pm 8.7)/\sqrt{n_{\rm ISM}}$ \ks, where n$_{\rm ISM}$ is the number density of the local ISM.
}

\maketitle
%
%________________________________________________________________

\section{Introduction}

Ever since the discovery paper by Becklin et al. (1969), the object
IRC +10 216 (= AFGL 1381 = CW Leo) has spurred much interest. We
now know that it is a carbon star in an advanced stage of stellar
evolution on the asymptotic giant branch (AGB), pulsating and surrounded by an optically
thick dust shell and large molecular circumstellar envelope (CSE).
One aspect of study has been to constrain the properties of the CSE by answering questions such as what is the mass-loss rate and how has it changed with time, what
kind of chemistry takes place, and what is the geometry and structure of the CSE?

The deep optical images taken by 
%Crabtree et al. (1987), 
Mauron \& Huggins (1999, 2000), 
Mauron et al. (2003), and Le\~{a}o et al. (2006)
show that the dusty envelope is not smooth but consists of a series of
arcs or incomplete shells. The average angular separation between the
dust arcs suggests a timescale for the change in mass-loss rate of the
order of 200--800 yr.  The lack of kinematic information on the dust
arcs precludes any firm conclusion about the true three-dimensional
structure of the arcs or shells.

From large-scale mapping at a relatively low angular resolution of the
CO J=1-0 emission, 
%from the envelope of CW Leo, 
Fong et al. (2003)
discovered a series of large molecular arcs or shells at radii of
$\sim$100\arcsec\ in the outer envelope. They attribute these multiple shells as 
``being the reverberations of a single Thermal Pulse erupting over 6000 yr ago.''
The timescale inferred from the spacing between these arcs is
about 200--1000 yr. 
%In addition, they suggest that the dust arcs seen
%in optical images are actually projections on the plane of the sky of
%the molecular arcs observed in CO, even though the dust arcs
%are found much closer ($\sim$20\arcsec--60\arcsec) to the central star.

%
In the present paper, we discuss our new results on the outer
shell of CW Leo from observations with the {\em Herschel Space Observatory} 
(Pilbratt et al. 2010) and their connection to the results from
{\em GALaxy Evolution Explorer Space Observatory} (GALEX) by Sahai \&
Chronopoulos (2010, hereafter SC).
For the present analysis, we adopted a distance of $d=$ 135\,pc and 
a mass-loss rate \mdot= 2.2$\cdot 10^{-5}$ \msolyr\ (Groenewegen et al. 1998), a gas expansion velocity of
V$_{\rm exp}$ = 15.4 \ks, a radial velocity V$_{\rm LSR}= -25.5$ \ks\ (Groenewegen et al. 2002), 
corresponding to V$_{\rm helio}= -18.6$\,\ks,
and a proper motion (pm) 
$\mu_{\rm \alpha} \cos \delta = +26 \pm$6, $\mu_{\rm \delta} =+4 \pm$6 mas/yr (Menten et al. 2006).
%\vspace{-5mm}

\section{Observations and data reduction}
\label{obsandreduc}

%\subsection{Observations}

The observations were carried out using the Photo Array Camera \&
Spectrometer (PACS, Poglitsch et al. 2010) and the Spectral and
Photometric Imaging REceiver (SPIRE, Griffin et al. 2010), 
and are part of the Science Demonstration Phase observations of the Mass-loss of Evolved StarS (MESS)
guaranteed time key program (Groenewegen et. al. 2010, in prep.), which is
investigating the dust and gas chemistry and the properties of CSEs
around a large sample of post-main-sequence objects.

For both instruments, the scan--map observing mode was used for a total
sky coverage of 30$\arcmin \times$30$\arcmin$.  In this mode, the
telescope is slewed at constant speed (20$\arcsec$/s for the PACS
data, 30$\arcsec$/s for the SPIRE data) along parallel lines to cover
the required area of the sky.  For the PACS observations, two scan maps
were taken with a scanning angle of 90$^\circ$ between the two to
achieve the most homogeneous coverage. The observation identification
numbers (obs.ID) for the two scans are: 1342186298 and 1342186299, with
a total integration time of 8.35 hours.  For the SPIRE observations, a
single scan was taken (obs.ID 1342186293) with a total integration
time of 1.35 hours. All observations were taken on 25 Oct. 2009.

\begin{figure*}[t]
\begin{center}
\vspace{0cm}
\hspace{0cm}
\resizebox{16.2cm}{!}{ \includegraphics[angle=0]{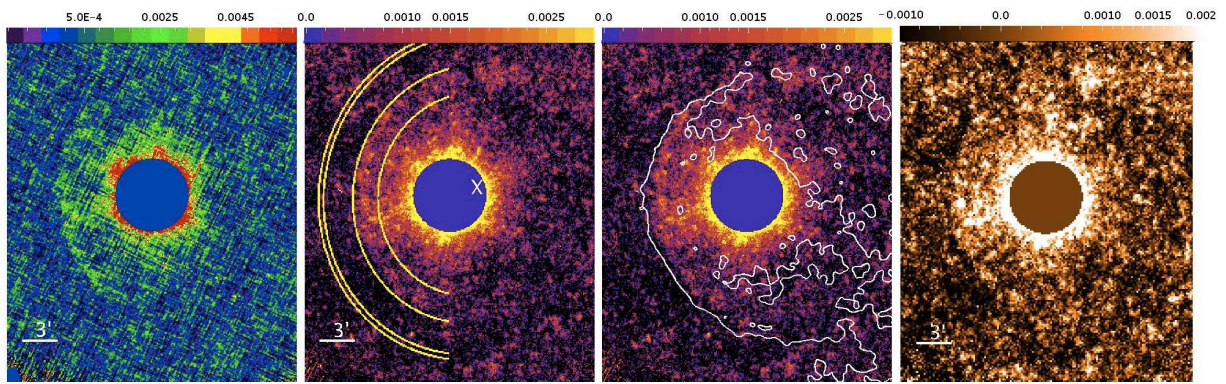}}
\caption{Surface brightness maps in Jy/pixel.
From left to right :
PACS 160~$\mu$m image.
SPIRE 250~$\mu$m image for which we overplotted the annuli segments between 
which we integrated the flux for the extended emission (inner annulus) and the sky (outer annulus).
The white cross represents the center of the ellipse.
SPIRE 250~$\mu$m image with overplotted in white the contour from the FUV map at 4.4$\times$10$^{-5}$ mJy$\cdot$arcsec$^{-2}$ limit.
SPIRE 350~$\mu$m image.
The FOV for all images is 23$\arcmin \times$27$\arcmin$. 
Background sources were removed from all maps.
% using HIPEs Daophot
%source extraction routine (using HIPE development version 3.0).
%
North is up and east is to the left.}
\label{red_PSW}
\vspace{-4mm}
\end{center}
\end{figure*}

%\subsection{Data reduction}

The data reduction was performed using the Herschel Interactive
Processing Environment (HIPE)
%\footnote{http://herschel.esac.esa.int/HIPE\_download.shtml} 
version 2.3. 
We followed the basic data reduction steps as described in the PACS data reduction guide
v1.2
%\footnote{http://herschel.esac.esa.int/Docs/DP/PACS\_Data\_Reduction\_Guide.pdf}
and the SPIRE data users manual
v1.0
%\footnote{http://herschel.esac.esa.int/Docs/DP/SPIRE\_dum.pdf}
with the exception of the median filtering of the background. 
In the SPIRE pipeline, the entire scanline is used to calculate the median value. 
In the PACS pipeline, a running median filter is used with a certain ``high pass filter width''.
%on a spatial scale of 4$\cdot$HPFW arcseconds (for 20\arcsec/s scanspeed).
%
A value of 450 was adopted that leads to a spatial scale of filtering
approximately equal to that used for the SPIRE data, and is of the order of 30\arcmin.
For both PACS and SPIRE, the central star was masked before applying
the median filter. The mask was a circle around the central star that
approximated a contour at 3 $\sigma$ of the background noise level.

Near-UV (NUV) and far-UV (FUV) GALEX data (Martin et al. 2005) were added to the analysis.
The data were taken in 2008 and are centred around 1528\AA\ (FUV) and 2271\AA\ (NUV). 
We used the pipeline product retrieved from the GALEX archive.
%\footnote{ http://galex.stsci.edu/GR4/}.

\vspace{-3mm}

\section{Analysis}
\label{analysis}

In the present paper, we do not comment on any structure close to the
star previously detected in the optical and in CO. No similar structure
is clearly visible in our data but a detailed analysis of the central structure requires a very
detailed understanding and accurate subtraction of, or deconvolution
with, the complicated PSF. This is beyond the scope of this paper.

The new result presented here is the extended emission
in the form of an arc clearly seen at 160~$\mu$m, 250~$\mu$m, and 350~$\mu$m 
with a spatial scale as large as 22$\arcmin$ (see Fig.~\ref{red_PSW}).
The arc is eastward of the central star with an almost circular curve. 
The central star is slightly to the east of the centre of the
structure and the arc is slightly flattened in the easterly direction. 
It is interesting to note that this extended emission matches the position and the shape of the FUV
extended emission (see Fig.~\ref{red_PSW}). In the Herschel
images, we do not see the patches of nebulosity seen in the western
part of the FUV image (see Fig.~1 in SC).

\begin{figure}
\begin{center}
\vspace{0cm}
\hspace{0cm}
\resizebox{9cm}{!}{ \includegraphics[angle=90]{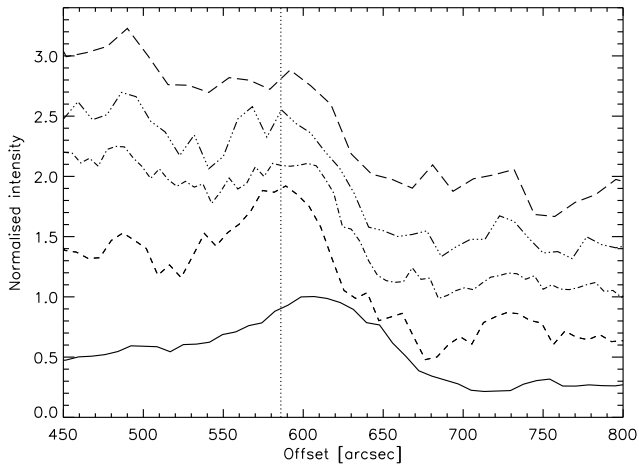}}
\caption{Intensity profiles as a function of the offset along the minor axis of the ellipse increasing from west to east. 
The profile for each wavelength was normalised to the intensity at 558$\arcsec$ and shifted up for clarity. From
bottom to top with normalisation factor in mJy/arcsec$^2$ and the vertical shift: FUV (0.065,0.0), 160~$\mu$m (0.063,0.8), 250~$\mu$m (0.023,1.1), 350~$\mu$m (0.008,1.4), 500~$\mu$m (0.004,1.8). 
The vertical dotted line indicates the position of the dust emission peak from the center of the ellipse.}
\label{Ipro}
\vspace{-4mm}
\end{center}
\end{figure}

The 1D intensity profile for each map (Fig.~\ref{Ipro}) was
constructed by subtracting the total flux integrated within two
successive apertures. To match the shape of the arc,
elliptical aperture photometry was used. We only considered the area
of the ellipse between 14\degr\ position angle (PA) and 167\degr\ PA
(see Fig.~\ref{red_PSW}) to match the spatial scale of the extended
emission. The ellipses are centred on $\alpha$\,=\,146.955$^\circ$ and
$\delta$\,= \,+13.28$^\circ$ with a PA of $-$12\degr\ and a
major-to-minor axis ratio taken to be 1.1.
For the emission from the sky, the inner minor axis is 13.3$\arcmin$ and the outer minor axis is 13.8$\arcmin$ (see Fig~\ref{red_PSW}).
The PACS fluxes were corrected following the numbers in the PACS
Scan Map release note 
%(PICC-ME-TN-035, 23 Feb 2010) 
yielding an absolute flux calibration uncertainty of 10--20\%. A
similar correction was applied to the SPIRE fluxes following the
correction factors given by Griffin et al. (2010) and Swinyard et
al. (2010). The SPIRE absolute flux calibration uncertainty is $\sim$15\%.
The dust shell is detected at 160~$\mu$m, 250~$\mu$m, and 350~$\mu$m
around 9.6$\arcmin$ from the star for a maximal surface brightness of
0.067~mJy$\cdot$arcsec$^{-2}$ at 160~$\mu$m. There is an offset of
$\sim$20$\arcsec$ between the FUV intensity peak and the far-IR
intensity peaks, which suggests that they have different origins.

\begin{figure}[t]
\begin{center}
\vspace{0cm}
\hspace{0cm}
\resizebox{9cm}{!}{ \includegraphics[angle=90]{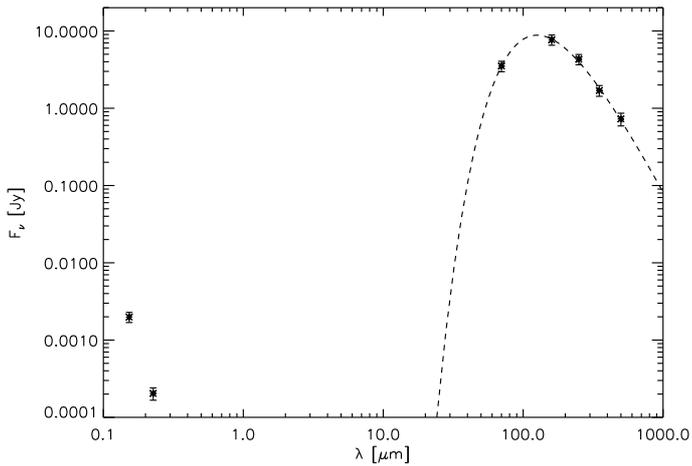}}
\caption{Modified black body ($T$= 25$\pm$3~K and $\beta$= 1.6$\pm$0.4) fit (dashed curve) to the derived fluxes (symbols) 
for the extended emission. }
\label{dustT}
\vspace{-4mm}
\end{center}
\end{figure}

Based on the location of the extended emission at 160~$\mu$m, the total flux in the arc was
calculated within the segment of an elliptical annulus of 150$\arcsec$ width, and inner
minor axis of 8.5$\arcmin$ between PA 14\degr\ and 167\degr\ (see Fig~\ref{red_PSW}). 
The derived fluxes are listed in Table~\ref{TableF}. 
The errors include the absolute calibration error and the error in the background estimation.
We fitted the photometry of the
extended emission with a function of the form $B_{\nu} \cdot \lambda^{-\beta}$, 
expected for a grain emissivity as $Q_{\rm abs} \sim \lambda^{-\beta}$. 
The NUV and FUV data are plotted for reference, but these do not fit
because the emission has a different physical origin (see below).
The best fit solution infers a dust temperature of 25$\pm$3~K and $\beta =$
1.6$\pm$0.4 (see Fig.~\ref{dustT}).  The value for $\beta$ is
in--between that expected for amorphous carbon (Rouleau \& Martin 1991,
$\beta$= 1.1) and astronomical silicates (Volk \& Kwok 1988, $\beta$=
2.0), suggesting that the material in the bow shock is a mixture of
C-rich material, lost by the star, and swept-up ISM material.

\section{Discussion}
\label{discussion}

\subsection{Bow shock, thermal pulse, or both ?}

The most widely accepted explanation of large detached shells is mass--loss 
variation (e.g., Olofsson et al. 1990, Zijlstra et al. 1992).
AGB stars experience thermal pulses (TPs) during which
intense mass loss ejections occur. A star can undergo several TPs, which would 
lead to the formation of concentric spherical detached shells (see Kerschbaum et al. 2010).

Another explanation of detached envelopes is the interaction between
the AGB wind and the interstellar medium (ISM) (e.g. Young et al. 1993, 
Martin et al. 2007, Ueta et al. 2006).  In this scenario, the AGB wind
is slowed down as it sweeps up material from the ISM. The piled up
material forms a density enhancement that continue to expand due to the
internal pressure. Shocks can occur if the relative velocity of the AGB wind
with respect to the ISM is large enough. The thermal emission of the dust in the density
enhancement at the shock interface between the stellar wind and the
the ISM can be detected in the far-IR (Ueta et al. 2006, 2009).  
While a detached shell produced by TPs would be spherical, in the case
of wind-ISM interaction the shape of the shell will depend on the
space motion of the star through the ISM. A wind--ISM shell can look
spherical as seen for R Cas (Ueta et al. 2009) if most of the space motion 
of the star relative to the ISM is in the radial direction. 
For stars with a high space motion
(relative to the ISM), the shape of the bow shock will be more
parabolic with the apex of the parabola in the direction of the star's
motion relative to the ISM.  In the case of CW\,Leo, the shape of the extended
emission and the position of the star suggest that the stellar wind
has driven a shock into the ISM. The far-IR emission is probably caused
by the thermal emission of the piled-up dust at the shock interface.

One may think that the observed UV emission is produced by dust
scattering of the interstellar light.
However, the brightness ratio of FUV to NUV is $\sim$10 
(see Table~\ref{TableF}), which is much higher 
than expected in that case (of the order of $\sim$2.4; SC). 
%This is supported by the fact that it is not possible
%to fit both the derived fluxes for the dust shell at the PACS and
%SPIRE wavelengths and the FUV and NUV fluxes using one black body
%temperature (see Fig.~\ref{dustT}).  
The only other AGB star with UV data probing a wind--ISM interaction
is Mira (Martin et al. 2007).  Martin et al. suggest that collisional
excitation of cool H$_2$ by hot electrons from the post-shock gas is
responsible for the UV emission. The faint NUV emission can be
explained by the H$_2$ emitting in the FUV band. SC suggest that the same mechanism 
may also be the dominant contributor to the FUV
ring emission in CW Leo.
The dust and the FUV emission have a similar spatial scale.
For the planetary nebula NGC 6720, we note that van Hoof et al. (2010) find
on the basis of PACS and SPIRE data  that dust and H$_2$ are
co-spatial and argue that H$_2$ has been formed on grain surfaces.

The structure seen in our data may also be the result of both mechanisms
i.e., mass--loss variation and wind--ISM interaction.  
By estimating the interpulse period for CW Leo, we can check wether
the dust shell we see is from an earlier TP that is now
interacting with the ISM.

Guelin et al. (1995) constrained the initial mass of CW Leo to be 3
\less\ $M$ \less\ 5 \msol\, based on the isotopic ratios of $^{24,25,26}$Mg.
From the initial--final mass relation from Salaris et al. (2009), this
implies a likely final mass (and essentially the current core mass) of
0.7--0.9 \msol.
From the core mass interpulse period relation of Wagenhuber \& Groenewegen (1998) 
for solar metallicity stars, this implies an interpulse period of  6~000--33~000 years.

%log ip = (-3.628 + 0.1337zi)(Mc - 1:9454)

Scaling the mass loss and the wind velocity in SC with our assumed
values, the flow timescale in the unshocked and shocked winds is
19~900 and 56~000 yr (very long because of the very low velocity of
1.2~\ks\ in this region, see SC for details), for a total lower limit
to the duration of the mass-loss phase of about 75~000 years.
%
%For a constant mass-loss rate this corresponds to a total mass in the envelope of $>$1.7 \msol.  
%
If TPs were to modulate the mass loss, as hypothesised for the origin for the
detached shells discussed earlier, one could expect at least one other
TP to have occurred during the time it took the envelope to expand
this far. No obvious density enhancement is evident in the unshocked
wind in the PACS and SPIRE images, which suggests that the interpulse
period is at least 19~000 years.
However, the dynamical and interpulse timescales are compatible, so it
is possuble that the bow shock and dust emission are not only the
result of a steady outflow interacting with the ISM, but might include
the effect of an enhanced wind of short duration
due to a recent TP (or pulses).

\begin{table}[!h]
\caption{Derived fluxes for the extended emission}
\begin{tabular}{cccccc}
\hline
\hline
$\lambda$  &  Flux & $\lambda$ &  Flux & $\lambda$ &  Flux    \\
($\mu$m)   &  (Jy) & ($\mu$m)   &  (Jy) & ($\mu$m)   &  (Jy) \\
\hline
0.15 & $(20\pm0.2)\times10^{-4}$  &  70  & 3.51$\pm$0.54 &  250 & 4.31$\pm$0.66 \\
0.23 & $ (2\pm0.2)\times10^{-4}$ & 160  & 7.70$\pm$1.17 &  350 & 1.70$\pm$0.27 \\
    &                          &      &               &  550 & 0.73$\pm$0.14 \\
\hline
\end{tabular}
\label{TableF}
\vspace{-7mm}
\end{table}

\subsection{Space motion of the star and the ISM flow velocity}

From the adopted radial velocity, distance, and proper motion, the
heliocentric space velocity of the star is derived to be about 25.1 \ks\
at a heliocentric inclination angle of the space motion vector of
47.8 degrees (measured from the plane of the sky, away from us)
and a PA of 81.3\degr\ for the proper-motion vector in the plane of the sky.

In the PACS/SPIRE wavebands, the observed far-IR surface brightness is
expected to be proportional to the column density of the dusty
material in the shell because the optical depth of the shell is much
lower than unity.  Because the bow shock interface is a parabolic
surface arbitrarily oriented in space, the column density tends to
reach its highest value where the bow shock cone intersects with the plane
of the sky including the central star.  Thus, the apparent shape of
the bow shock is the conic section of the bow shock.  Therefore, given
the analytical formula of the bow shock structure (Wilkin 1996), we
can fit the observed surface brightness of the bow shock to the conic
section of the bow shock cone to derive the heliocentric orientation
of the bow shock (e.g. Ueta et al. 2008; 2009).  From this fitting,
we determined that the apex of the bow shock cone is oriented at
$61\fdg9 \pm 0\fdg3$ (this is degenerate, in a sense that it could be
pointing away from us or towards us) with respect to the plane of the
sky into the PA of $88^{\circ}$ with the deprojected stand-off
distance of $(8.0 \pm 0.3) \times 10^{17}$ cm.

% Eq 2 in Ueta 0911.4918v1 R0= 4.65 e21 sqrt( Mdot[msol/yr] * v [km/s] / n_ISM * (V_*[km/s])^2 )
Using these numbers in the ram pressure balance equation 
$V_{\star} = \sqrt{\frac{\dot{M} \; V_{\rm exp}}{4 \pi \; \mu_{\rm H} \; m_{\rm H} \; n_{\rm ISM} \; R_0^2} }$
for a mean nucleus number per hydrogen nucleus of $\mu_{\rm H}= 1.4$, the relative velocity of the star with respect to the ISM is
$V_{\star} = (106.6 \pm 8.7)/\sqrt{n_{\rm ISM}}$ \ks,
where $n_{\rm ISM}$ is the number density of the ISM local to CW Leo in cm$^{-3}$.

By following the scheme of Johnson \& Soderblom (1987, also see Ueta et al. 2008),  
the heliocentric space velocity components of the star can be converted to the 
heliocentric Galactic space velocity components [U, V, W] of
[21.6$\pm$3.9, 12.6$\pm$3.5, -1.8$\pm$3.3] \ks\ ,
% (magnitude 25.1 km/s)
and also to the LSR Galactic space velocity components of
[30.6$\pm$3.9, 24.6$\pm$3.5, 5.2$\pm$3.3] \ks.
%(magnitude 39.6 km/s)
%
The heliocentric ISM flow velocity is 
117.6 \ks\ if the bow cone is facing us (i.e.~the apex pointing toward)
or
82.6 \ks\ if the bow cone is facing away from us.
%(i.e.~the apex is pointing away from us).

Wareing et al. successfully modelled bow shocks for AGB stars using
3-D hydrodynamical models (R\,Hya and Mira in Wareing et al. 2006,
2007a). Their models were able to reproduce in great detail all the components
of a bow shock (astropause, astrotail, vortices) and constrain the
space velocity of the star, the ISM density, and the mass--loss rate.
The models were built by varying the velocity of the star
relative to the ISM, the ISM density and the mass loss rate and the
effect of those physical parameters on the shape of the bow shock can
be seen in Wareing et al. (2007b). From the Wareing models, we can see
that for ISM densities $\le$2~cm$^{-3}$ (which implies $V_{\star}
\ge$75 \ks) and after an evolution of 50~000~yr into the AGB phase,
the bow shock does not resemble our data. The models
have a more flattened bow shock with the star closer to the apex of
the shock. This is for a mass--loss rate between 5$\times
10^{-7}$~\msolyr\ and 10$^{-6}$~\msolyr, which is an order of
magnitude lower than the value assumed for CW\,Leo. For a higher mass--loss
rate, we would expect an even more important departure from
sphericity with a significant tail of ejecta.  This suggests that the
density of the local ISM to CW\,Leo is probably higher than
2~cm$^{-3}$ implying an upper limit of 75 \ks to $V_{\star}$, the stellar velocity relative to the ISM.

\begin{acknowledgements}
D.L, M.G, J.B, W.D, K.E, R.H, C.J, S.R, P.R and B.V acknowledge support from 
the Belgian Federal Science Policy Office via the PRODEX Programme of ESA.
PACS has been developed by a consortium of institutes led by MPE
(Germany) and including UVIE (Austria); KUL, CSL, IMEC (Belgium); CEA,
LAM (France); MPIA (Germany); IFSI, OAP/AOT, OAA/CAISMI, LENS, SISSA
(Italy); IAC (Spain). This development has been supported by the funding
agencies BMVIT (Austria), ESA-PRODEX (Belgium), CEA/CNES (France),
DLR (Germany), ASI (Italy), and CICT/MCT (Spain).
SPIRE has been developed by a consortium of institutes led by
Cardiff Univ. (UK) and including Univ. Lethbridge (Canada);
NAOC (China); CEA, LAM (France); IFSI, Univ. Padua (Italy);
IAC (Spain); Stockholm Observatory (Sweden); Imperial College London,
RAL, UCL-MSSL, UKATC, Univ. Sussex (UK); and Caltech, JPL, NHSC,
Univ. Colorado (USA). This development has been supported by
national funding agencies: CSA (Canada); NAOC (China); CEA,
CNES, CNRS (France); ASI (Italy); MCINN (Spain); SNSB
(Sweden); STFC (UK); and NASA (USA)
Data presented in this paper were analysed using ``HIPE'', a joint
development by the Herschel Science Ground Segment Consortium,
consisting of ESA, the NASA Herschel Science Center, and the HIFI,
PACS and SPIRE consortia.
FK acknowledges funding by the Austrian Science Fund FWF under project
numbers P18939-N16 and I163-N16.

\end{acknowledgements}

\clearpage
\newpage
\normalsize
\begin{appendix}

\end{appendix}


\begin{thebibliography}{}
%\footnotesize
%\scriptsize

\bibitem[]{} Becklin, E.E., Frogel, J.A., Hyland, A.R., Kristian, J., Neugebauer, G. 1969, ApJ, 158, L133

%\bibitem[]{} Crabtree, D.R., McLaren, R.A., \& Christian, C.A. 1987, in Late Stages of
%Stellar Evolution, ed. S. Kwok, S.R. Pottasch, ASSL, 132, 145

\bibitem[]{} Griffin, M.J., Abergel, A., Ade, P.A.R., et al. 2010, A\&A this issue

\bibitem[]{} Fong, D., Meixner, M., Shah, R.Y. 2003, ApJ, 582, L39

\bibitem[]{} Groenewegen, M.A.T., van der Veen, W.E.C.J. \& Matthews, H.E. 1998, A\&A, 339, 489

\bibitem[]{} Groenewegen, M.A.T., Sevenster, M., Spoon, H.W.W., \& Perez I. 2002, A\&A, 390, 501 

\bibitem[]{} Guelin, M., Forestini, M., Valiron, P.,  et al. 1995, A\&A, 297, 183

\bibitem[]{} Johnson, D. R. H., \& Soderblom, D. R. 1987, AJ, 93, 864

\bibitem[]{} Kerschbaum, F., Ladjal, D., Ottensamer, R., et al. 2010, A\&A, this issue

\bibitem[]{} Le\~{a}o, I.C., de Laverny, P., M\'ekarnia, D., de Medeiros, J.R., \& Vandame, B. 2006, A\&A, 455, 187

%\bibitem[]{} Luntilla, T. \& Juvela, M. 2007, A\&A, 470, 259

\bibitem[]{} Martin, D.C., Fanson, J., Schiminovich, D., et al. 2005, ApJ, 619, L1

\bibitem[]{} Martin, D.C., Seibert, M., Neill, J.D., et al. 2007, Nature, 448, 780

\bibitem[]{} Mauron, N., \& Huggins, P.J. 1999, A\&A, 349, 203

\bibitem[]{} Mauron, N., \& Huggins, P.J. 2000, A\&A, 359, 707

\bibitem[]{} Mauron, N., de Laverny, P., \&  Lopez, B. 2003, A\&A, 401, 985

\bibitem[]{} Menten, K.M., Reid, M.J.,  Kr\"ugel, E.,  Claussen, M.J. \& Sahai, R. 2006, A\&A, 453, 301

%\bibitem[]{} Morrissey, P., Schiminovich, D., Barlow, T.A., et al. 2005, ApJ, 619, 7

%\bibitem[]{} Mezger, P.G., Mathis, J.S., \& Panagia, N. 1982, A\&A, 105, 372

\bibitem[]{} Olofsson, H., Carlstr\"{o}m, U., Eriksson, K., et al. 1990, A\&A, 230, L13

\bibitem[]{} Pilbratt, G., et al. 2010, A\&A this issue

\bibitem[]{} Poglitsch, A., Waelkens, C., Geis, N., et al. 2010, A\&A this issue

\bibitem[]{} Rouleau, F., \& Martin, P.G., 1991, ApJ, 377, 526 

\bibitem[]{} Sahai, R., \& Chronopoulos, C.K. 2010, ApJ, 711, L53

\bibitem[]{} Salaris, M., Serenelli, A., Weiss, A., Miller Bertolami, M. 2009, ApJ, 692, 1013

\bibitem[]{} Swinyard, B.M., Ade, P.A.R., Baluteau, J.-P, et al. 2010, A\&A this issue

\bibitem[]{} Ueta, T., Izumiura, H., Yamamura, I., et al. 2008, PASJ, 60, S407

\bibitem[]{} Ueta, T., Hideyuki, I., Yamamura, I., et al. 2009, arXiv0905.0756

\bibitem[]{} Ueta, T., Speck, A.K., Stencel, R.E. et al. 2006, ApJ, 648, L39

%\bibitem[]{} van Buren, D. \& McCray, R. 1988, ApJ, 329, L93

\bibitem[]{} van Hoof, P.A.M., Barlow, M.J., Van de Steene, G.C., et al. 2010,  A\&A this issue

\bibitem[]{} Volk, K., \& Kwok, S. 1988, ApJ, 331, 435

\bibitem[]{} Wagenhuber, J. \& Groenewegen, M.A.T. 1998, A\&A, 340, 183

\bibitem[]{} Wareing, C.J., Zijlstra, A.A., Speck, A.K. et al. 2006, MNRAS, 372, L63

\bibitem[]{} Wareing, C.J., Zijlstra, A.A., \& O'Brien, T.J. 2007b, MNRAS, 382, 1233

\bibitem[]{} Wareing, C.J., Zijlstra, A.A., \& O'Brien, T.J., \& Seibert, M. 2007a, ApJ, 670, L125

\bibitem[]{} Wilkin, F. P. 1996, ApJ, 459, L31

\bibitem[]{} Young, K., Phillips, T.G., Knapp, G.R. 1993, ApJ, 409, 725

\bibitem[]{} Zijlstra, A.A., Loup, C., Waters, L.B.F.M., de Jong, T. 1992, A\&A, 265, L5


\end{thebibliography}
\end{document}